\documentclass[twoside]{article}
\usepackage{Proc_NTSE}

\begin{document}
\thispagestyle{plain}

\begin{center}
{\Large \bf \strut
 Three-Nucleon Forces Revisited - \\ Some Historical Thoughts
\strut}\\
\vspace{10mm}
{\large \bf Peter U. Sauer}
\end{center}

\noindent{\small \it Institute for Theoretical Physics, Leibniz University, D-30167 Hannover, Germany} \\
\markboth{Peter U. Sauer}
{Three-Nucleon Forces} 

\begin{abstract}
Historic steps in the emergence, the derivation and the use of three-nucleon forces, genuine and effective, for calculations of few-nucleon systems and of the structure of heavier nuclei are recalled. The research focus is on few-nucleon systems. The need of three-nucleon forces for a successful description of some data and the remaining puzzles of other data, not explainable despite the inclusion of three-nucleon forces, are discussed.
\\[\baselineskip] 
{\bf Keywords:} {\it Nuclear forces; shell model; few-nucleon systems.}
\end{abstract}

\section{Introduction}

The shell-model theme of this conference is not my current research territory. I would not have attended, would the conference not also celebrate James Vary with whom I shared early stages of my carrier. I decided against a standard talk on actual research. Instead, I want first to  reflect on what drove our research then, before coming to the $\it Here \, and \, Now$, which is the nuclear shell model for James and few-nucleon systems for me. \\

I got to know James in 1970/71, when we were both postdocs in the nuclear theory group of MIT. We started  to collaborate on the challenge of that time, the derivation of nuclear properties from the interaction between free nucleons. And that challenge is still with us today, as this conference proves.

\section{My Personal View on the Nuclear Shell Model, Then and Now}

Doing microscopic nuclear structure in 1970/71, i.e., calculating the properties of nuclear matter, of doubly closed-shell nuclei and of simple shell-model systems  in terms of a realistic two-nucleon (2N) interaction, was a courageous enterprise: The suggested 2N potentials were scary beasts, their short-ranged core was conceptually unknown and, furthermore, it was parametrized in form of a  strong repulsion which had to be smoothened  into the in-medium reaction matrix of Brueckner theory \cite{brueckner:55}.\\  

\begin{figure}[!]
\centerline{\includegraphics[width=1.0\textwidth]{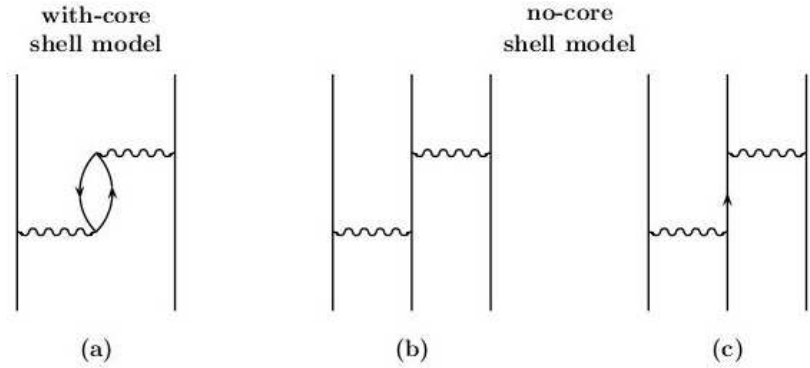}}
\caption{\label{fig:ShellModel} Strategies describing the contribution of core polarization to the effective shell-model interaction. Plot (a) represents the contribution in calculations with an inert core; plots (b) and (c) represent the same process when resolved in no-core calculations. The vertical lines without an arrow stand for nucleons in the model space, the horizontal wavy lines for the 2N Brueckner reaction matrix derived from the 2N potential. In plot (a) the backward arrow indicates a hole state in the inert core, the forward arrow a particle state outside the core; plot (a) is an irreducible 2N contribution to the effective interaction, irrespectively, if the particle state is within or outside the model space. However, if the particle state is inside the model space of a no-core calculation, the process is reducible into two subsequent interactions within the model space, as plot (b) shows. In contrast, if the particle state is outside the model space as in plot (c), the process remains irreducible within the model space and is part of an effective 3N contribution to  the shell-model interaction.}
\end{figure}

At that time, James's and my common nuclear-structure playground was the shell-model of $^{18} \rm O$, described by an inert $^{16} \rm O$ core with two active neutrons outside the core. The latter nucleons formed the active Hilbert space, the model space, consisting of 2s-1d states only, the corresponding effective interaction being the 2N reaction matrix, modified by core-polarization, shown in  Fig.~\ref{fig:ShellModel}(a); core polarization acts technically as an effective interaction between the two active nucleons, though, physically, it involves three nucleons. Kuo and Brown \cite{kuobrown:66} had initiated this game and appeared to have also closed the issue by their impressive achievement in describing data. But our revolutionary minds were challenged. We improved the calculation by better numerics \cite{vary:73} and found the numerical inadequacy of the effective interaction in use, therefore the distortion of our names Vary, Sauer and Wong in the author list by the community to {\em Very \, sorry, \, wrong!}; others \cite{barrett:68, schucan:72} challenged the whole shell-model strategy of that time on more fundamental grounds than we did. This was the dark moment of the early microscopic shell model. \\

Increasing computational capability of theoretical physics allowed a novel, more physical shell-model strategy, e.g., the description of  $^{18} \rm O$ without an inert $^{16} \rm O$ core \cite{zheng:93,navratil:96}: Use a model space, numerically manageable and physically large enough for accommodating the considered physics phenomena realistically, accompanied by a corresponding effective interaction, which should stay as simple as possible. This fact is illustrated in Figs.~\ref{fig:ShellModel}(b) and ~\ref{fig:ShellModel}(c) for the core-polarization contribution to the effective shell-model interaction in the no-core description. Of course, in the search for balance between model space and effective interaction the truncation of the full Hilbert space to the active model space generally remains necessary in shell-model calculations: The usually employed oscillator basis is advantageous for the symmetry and geometry of finite nuclei, but awkward when having to build up the tail behavior of single-particle states and when having to punch the correlation hole into the 2N wave function. Thus, the truncation of Hilbert space remains physically severe and makes effective many-body contributions to the interaction important. Even without {\em genuine} 3N forces, {\em effective} ones arise as from core polarization, shown in Fig.~\ref{fig:ShellModel}(c).  This search for an efficient balance between Hilbert space and interaction is a basic nuclear-structure problem also in a broader context outside the shell model; it is my theme throughout this talk. \\

At this special occasion, another paper with James and Pradhan of that early time \cite{pradhan:72} comes to my mind, a paper whose idea still echos in modern shell-model approaches:  The core region of the 2N force - now in meson theory the realm of omega- and rho-meson exchanges, in chiral effective field theory (EFT) the realm of two- and many-nucleon contact contributions - was for us {\em terra incognita } which we wanted to explore by the technique of short-ranged phase-equivalent off-shell variations, hoping to stumble on a novel, more pleasing parametrization of the 2N potential. In retrospect, we did not learn anything about that unknown part of the 2N force, since we were searching rather randomly in that paper. Our hope for information on the force from nuclear structure was a naive illusion at that time. But that hope is still behind the so-called {\em ab exitu} approach to the effective interaction \cite{shirokov:07} in no-core shell-model calculations, and it is still behind the modern and really clever use of phase-equivalent variations \cite{bogner:03}, in fact a smoothening procedure of the 2N potential -  a similar strategy as Brueckner theory used with its reaction matrix by the ladder summation of highly excited states -,  the prize to be payed being the rise of {\em effective} many-nucleon interactions even without a proper truncation of Hilbert space.  \\

The basic assumption of nuclear theory, before the advent of quantum chromodynamics and still now, is: {\em Rigid nucleons, the only active degrees of freedom in nuclei, interact through genuine two-, three- and possibly many-nucleon forces according to the rules of non-relativistic quantum mechanics.} That assumption confronts us with two distinct problems which in 1970/71 also defined different fields of research: First, assuming a parametrization of nuclear dynamics, how can we solve the many-nucleon problem throughout the period table? This is still the challenge for present-day {\em shell-model} calculations. But second, more basic, how can we learn details about those forces from some nuclear properties, if they are really reliably described theoretically? Our paper on phase-equivalent off-shell variations \cite{pradhan:72} mixed up both fields of research, and therefore hopelessly dealt with too complex problems. The second question is the field of {\em few-nucleon systems}. I chose that path of few-body physics for my later research which I  discuss next, but I shall remember, how my early research with James influenced what I am doing today. \\

\section{Few-Nucleon Systems}

The many-body problem is for few-nucleon bound and scattering states conceptually under control due to Faddeev \cite{faddeev:61} and Alt, Grassberger and Sandhas \cite{alt:67},  and it is getting, step-by-step, also calculationally under control by various numerical techniques. My collaborators and me adopted integral equations in momentum space as our numerical technique; compared to shell-model calculations of bound-state systems, the calculations are quite tricky for few-nucleon scattering due to singularities, though the singularities  are integrable; they arise from open inelastic channels. Results shown later on are obtained by that technique. The latest important technical achievements were the inclusion of the Coulomb interaction between protons (p) in the scattering equations \cite{deltuva:08a}, a stumbling block for the theoretical description during decades, and the description of 4N scattering above the four-particle breakup threshold \cite{deltuva:12}. On the experimental side, there is a multitude of data, especially now data of reactions with polarized particles. From those data one can hope to get more and more information on nuclear forces. I describe that project in its important steps.\\

\subsection{Choice of Dynamics}

\begin{figure}[!]
\centerline{\includegraphics[width=0.8\textwidth]{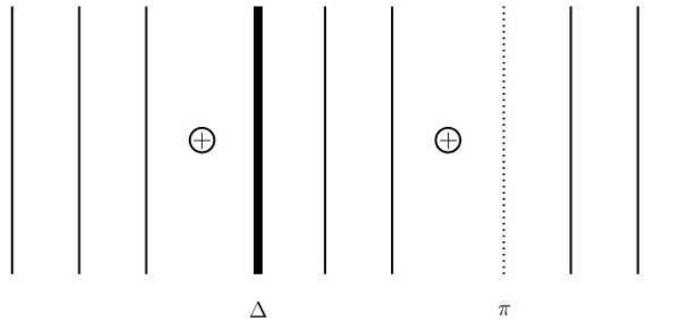}}
\caption{\label{fig:HilbertSpace} Hilbert space for the description of nuclear phenomena at low and intermediate energies. Compared with the purely nucleonic one, it is expanded by sectors, in which one N is turned into a $\Delta$ isobar and one $\pi$ is added to the N's. $\pi  \rm N$ scattering is described in the corresponding Hilbert space of baryon number one. The 2N reactions without and with a single $\pi$ are described in the corresponding Hilbert space of baryon number two.}
\end{figure}

The form of the nuclear dynamics to be tested has to be specified. We had to decide on our form, when pion factories were en vogue; the inclusion of pion production and absorption was necessary: Thus, the important active degrees of freedom to be considered were, besides the nucleon ($\rm N$), the pion ($\pi$) and the Delta-isobar ($\Delta$), which strongly mediates $\pi$ production in the 2N isospin-triplet partial waves; experimentally, the $\Delta$ isobar is observed as  ${\rm P}_{33}$ $\pi \rm N$ resonance;  single-$\pi$ production dominates well above 2$\pi$- and 3$\pi$-production thresholds. The chosen Hilbert space is shown in Fig.~\ref{fig:HilbertSpace}; in fact, the choice of an expanded Hilbert space is conceptually based on the same strategy which the no-core shell model took when including the physically important core degrees explicitly in the active model space: {\em Active degrees of freedom belong to the Hilbert space, they cannot be simulated well by a complicated hamiltonian.} That strategy \cite{sauer:86} allows a unified description of nuclear phenomena at low and at intermediate energies, e.g., the simultaneous description of 2N reactions, elastic and inelastic with single-$\pi$ production and absorption.\\

\begin{figure}[!]
\centerline{\includegraphics[width=0.8\textwidth]{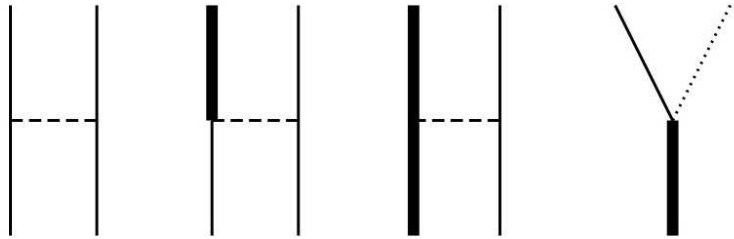}}
\caption{\label{fig:Hamilton} Hamiltonian describing the nuclear dynamics in the Hilbert space of Fig.~\ref{fig:HilbertSpace}. The interactions are of two-baryon nature, coupling purely nucleonic channels with those containing a $\Delta$ isobar; the latter ones are coupled to the pionic channels by a single-baryon vertex.}
\end{figure}

The hamiltonian corresponding to the chosen Hilbert space was taken from meson theory which was without alternative at that time. It is illustrated in Fig.~\ref{fig:Hamilton}, it consists of a one-baryon piece, mediating  $\pi \rm N$ scattering in the ${\rm P}_{33}$ partial waves - a $\pi \rm N$ potential is to be added for the non-resonant partial waves - and mediating $\pi$ production and absorption, and it consists of two-baryon potentials derived from all possible meson exchanges. That hamiltonian has a particular characteristic for the $\Delta$ isobar \cite{poepping:87}; it cannot be produced experimentally; the corresponding S-matrix element is exactly zero; observable are the coupled $\pi \rm N$ states. For that ambitious hamiltonian we were able to do calculations in most of its aspects \cite{sauer:86}, e.g., for all reactions in the two-baryon sector $ \rm NN \rightarrow NN$, $ \rm NN \rightarrow d \pi$, 
$ \rm NN \rightarrow NN \pi$, $ \rm d \pi \rightarrow d \pi$,  $ \rm d \pi \rightarrow NN \pi$ and $ \rm d \pi \rightarrow NN$  up to 0.5 GeV c.m. energy - $\rm d$ standing for the deuteron. But the hamiltonian was not well tuned to low-energy 2N data and therefore was not reliable enough for the description of few-nucleon systems at low energies, my more recent research focus. \\ 

\begin{figure}[h]
$\Large \rm \hspace{6mm} Fujita-Miyazawa \hspace{37mm} higher \,\,order \,\, 3N \,\, force \vspace{5mm}$ \\ 
\centerline{\includegraphics[width=0.9\textwidth]{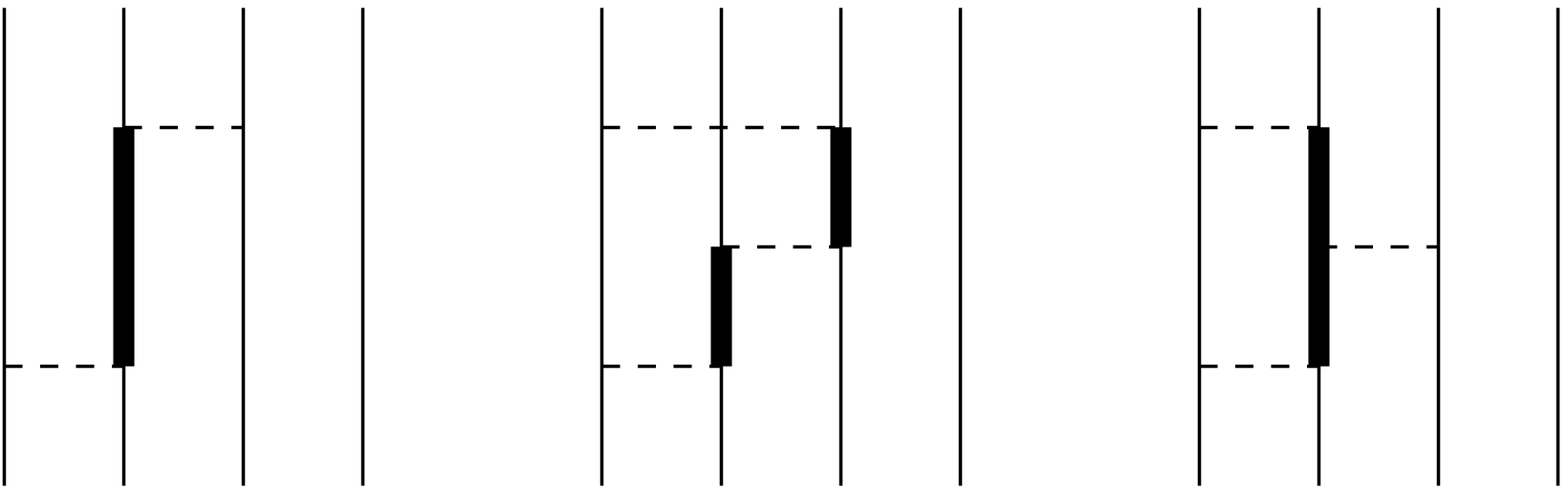}} \\  \\ \vspace{0mm}
$\Large \rm \hspace{58mm} 4N \,\, force \vspace{5mm} $ \\
\centerline{\includegraphics[width=0.6\textwidth]{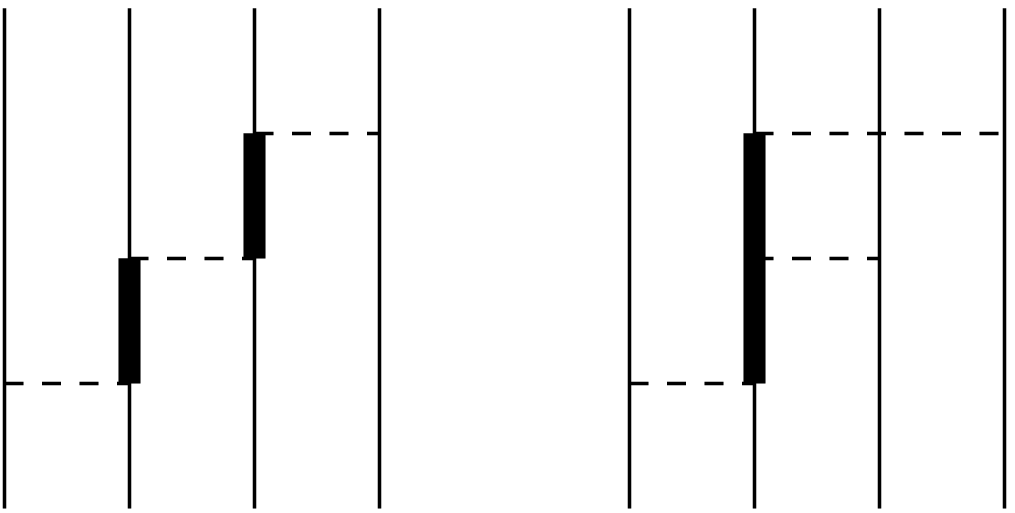}}
\caption{\label{fig:Forces} $\Delta$-mediated 3N and 4N forces, consistent with each other and with the 2N interaction. The upper row shows examples for the arising 3N force, the Fujita-Miyazawa process being the one of lowest order \cite{fumi:57}. The lower row shows examples for the arising 4N force. All possible meson exchanges are considered.}
\end{figure}

The explicit treatment of the $\Delta$ isobar has an important and wanted effect; it yields effective 3N, 4N and many-N forces; they are irreducible in the purely nucleonic Hilbert sector, but are resolved into two-baryon pieces in the expanded Hilbert space of Fig.~\ref{fig:HilbertSpace}. In standard meson theory and in standard EFT, 2N, 3N and many-N potentials arise from freezing non-nucleonic degrees of freedom; but vice versa, as done in the present approach, an important contribution to {\em genuine} 3N and many-N forces can be resolved, when keeping the $\Delta$-isobar degree of freedom explicitly.  And without active pions, i.e., without the one-baryon piece of Fig.~\ref{fig:Hamilton}, the hamiltonian is tuned well for the purposes of low energies \cite{deltuva:03}, i.e., below $\pi$-production threshold. The coupled 
two-baryon potential will be referred to as $\rm CD \, \rm Bonn + \Delta$; its purely nucleonic reference potential is $\rm CD \, \rm Bonn$, whose extension it is. Even that truncated hamiltonian provides consistent 2N, 3N and 4N forces, in general many-N forces, for what Fig.~\ref{fig:Forces} shows examples; their forms and strengths are fixed, they do not allow any further tuning to 3N and 4N data;  physicswise, those arising forces are still incomplete, since other mechanisms leading to irreducible many-N forces besides the $\Delta$-mechanism are left out. \\

I have discussed 3N and many-N forces from various angles. It is now appropriate to come to a conclusive summary: There are {\em genuine} and {\em effective} nuclear forces. \\

The {\em genuine} forces are derived in the form of instantaneous potentials of a many-N hamiltonian in a complete Hilbert space for the quantum-mechanical description of many-N systems; they incorporate accepted knowledge of the nuclear forces as the one-$\pi$ exchange tail; the remainder of the genuine 2N potential was phenomenological in the early days, was later on derived from meson theory and is now usually derived from EFT, i.e., from field theories with non-nucleonic degrees of freedom; in the step to the potential all non-nucleonic degrees are frozen; this step is non-unique. In the same way, 3N and many-N potentials are not made by God, they are babies of theoreticians and therefore in principle non-observable. When we loosely speak that some experimental data signal the dynamic need of a contribution from the 3N potential, we mean that in a chosen dynamic description the use of a 2N potential alone is insufficient. \\

The {\em effective} forces are by-products of particular solution techniques for the nuclear many-nucleon problem in the frame work of non-relativistic quantum mechanics. They arise when the complete Hilbert space has to be truncated, the arising 2N, 3N and many-N forces then correct for that truncation; those forces are often energy-dependent, i.e., time-delayed; the 2N reaction matrix of Brueckner theory is such an energy-dependent 2N force, it is also dependent on the amount of truncation. The {\em effective} forces also arise when the hamiltonian is transformed to act dominantly in a particular and convenient subspace, even without truncation, most conveniently in a subspace of low momenta; they are by-products of a particular smoothing technique. {\em Effective} 3N and many-N forces arise, even if the underlying hamiltonian consists of 2N {\em genuine} forces only. \\

In structure calculations of heavier nuclei {\em effective} many-N forces arise in the process of solving the nuclear many-body problem. In the description of few-nucleon systems at low and intermediate energies {\em genuine} many-N forces can be simulated as in Fig.~\ref{fig:Forces} by keeping non-nucleonic degrees of freedom explicitly in the active Hilbert space. \\

\subsection{Results for Few-Nucleon Bound States}

Hadronic and electromagnetic properties of ${^3} \rm H$, ${^3} \rm He$ and ${^4} \rm He$ are calculated. The effect of the 3N force on binding is sizable according to Ref.  \cite{deltuva:08b}, its Fujita-Miyazawa part \cite{fumi:57} being the dominant contribution, usually twice the other 3N-force contributions. In contrast, the effect of the 4N force on binding is small, in fact, an order of magnitude smaller than the 3N-force effect. This observation is the first solid confirmation of the general folklore on the hierarchy in many-N forces. Since the chosen dynamics cannot be tuned anymore, the resulting binding energies still fail the experimental values slightly. That miss of binding is therefore carried to the thresholds of reactions, a disadvantage for the description of 4N scattering close to thresholds. In contrast, the experimental binding-energy difference between ${^3} \rm H$ and ${^3} \rm He$ is well accounted for. \\

\subsection{Results for Few-Nucleon Reactions}

The few-nucleon community is able to account for a very large amount of experimental 3N and 4N data at low energies, i.e., at energies up to the $\pi$-production threshold. This is quite satisfying. The inclusion of Coulomb and of a 3N interaction is often needed; I give an example for both effects. Besides those successes which are in the overwhelming majority, there are, however, puzzles, i.e., there is a persistent disagreement between theoretical prediction and data without any hint for a solution; in fact that is the much more interesting situation, since we hope to learn from such cases; I shall also give an example for such a puzzle. In the presented figures, the predictions derived from the coupled-channel potential $\rm CD \, \rm Bonn + \Delta$ with Coulomb, indicated by $\Delta + \rm Coulomb$ and by the red curves, are the most complete ones, including the effect of Coulomb and of many-N forces mediated by the $\Delta$ isobar simultaneously. The predictions derived from the purely nucleonic reference potential $\rm CD  \,\rm Bonn $ with Coulomb, indicated by $\rm N + \rm Coulomb$ and by the green curves, include the effect of Coulomb, but leave out the effect of many-N forces mediated by the $\Delta$ isobar; the difference between red and green curves indicate the effect of many-N forces on the considered observable. The predictions derived from the coupled-channel potential $\rm CD \, \rm Bonn + \Delta$ without Coulomb, indicated by $\Delta$ and by blue curves, leave out the effect of Coulomb, but include the effect of many-N forces mediated by the $\Delta$ isobar; the difference between red and blue curves indicate the effect of Coulomb on the considered observable. \\ 

\begin{figure}[!]
\vspace{0mm}
\centerline{\includegraphics[width=0.60\textwidth]{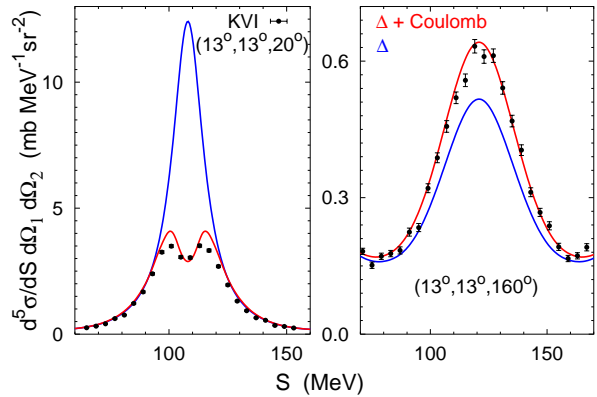}} 
\caption{\label{fig:Coulomb} dp breakup at 130 MeV d energy. The Coulomb effect is quite pronounced due to the correlation between the two protons in the final state. The angles of the two outgoing p's are fixed; their energies are constrained by the kinematical locus $\rm S$. There is no evidence for the need of a 3N force. The experimental data and the theoretical predictions are from Ref. \cite{kistryn:06}.}
\end{figure}

\begin{figure}[!]
\vspace{0mm}
\centerline{\includegraphics[width=0.40\textwidth]{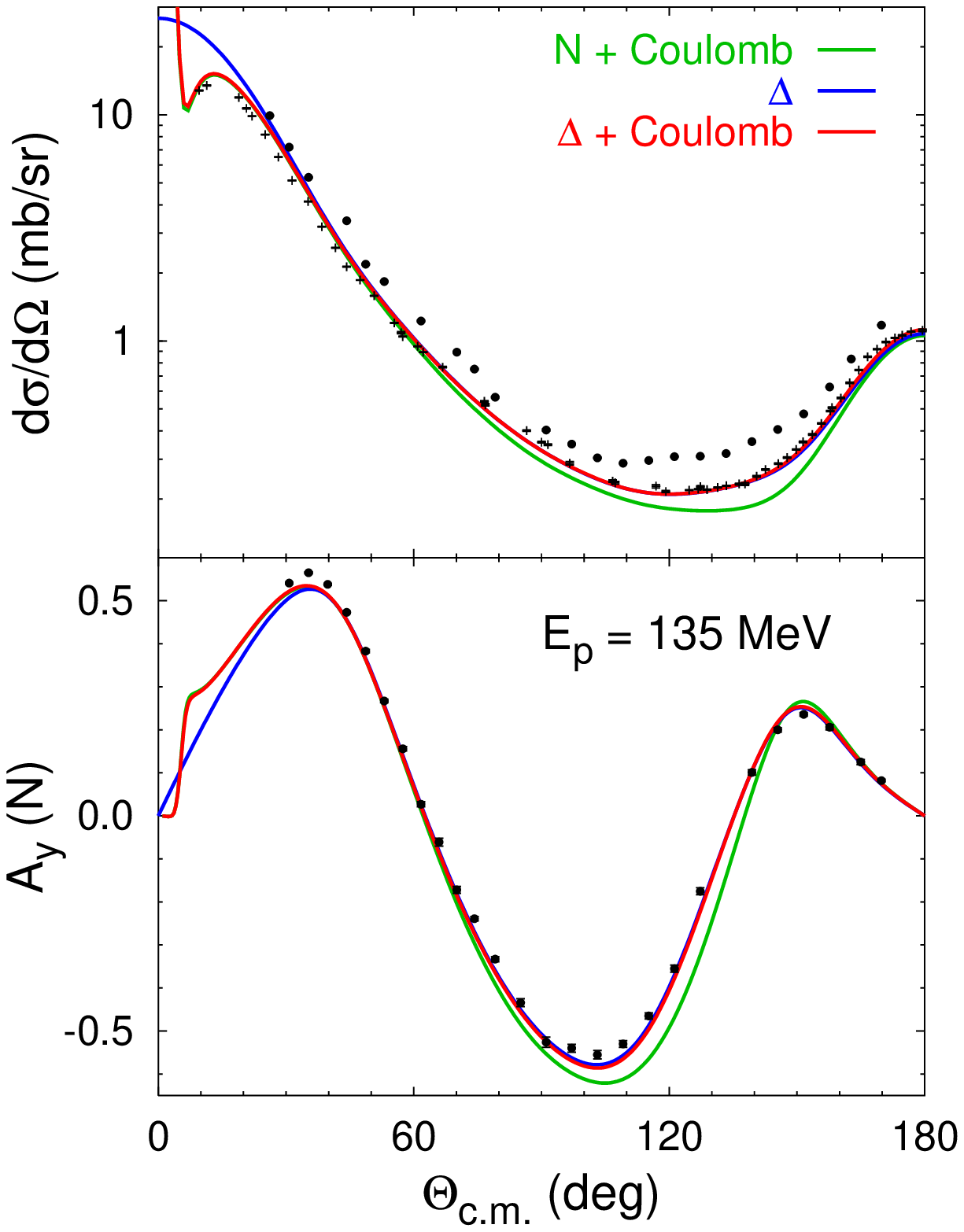} 
                    \includegraphics[width=0.40\textwidth]{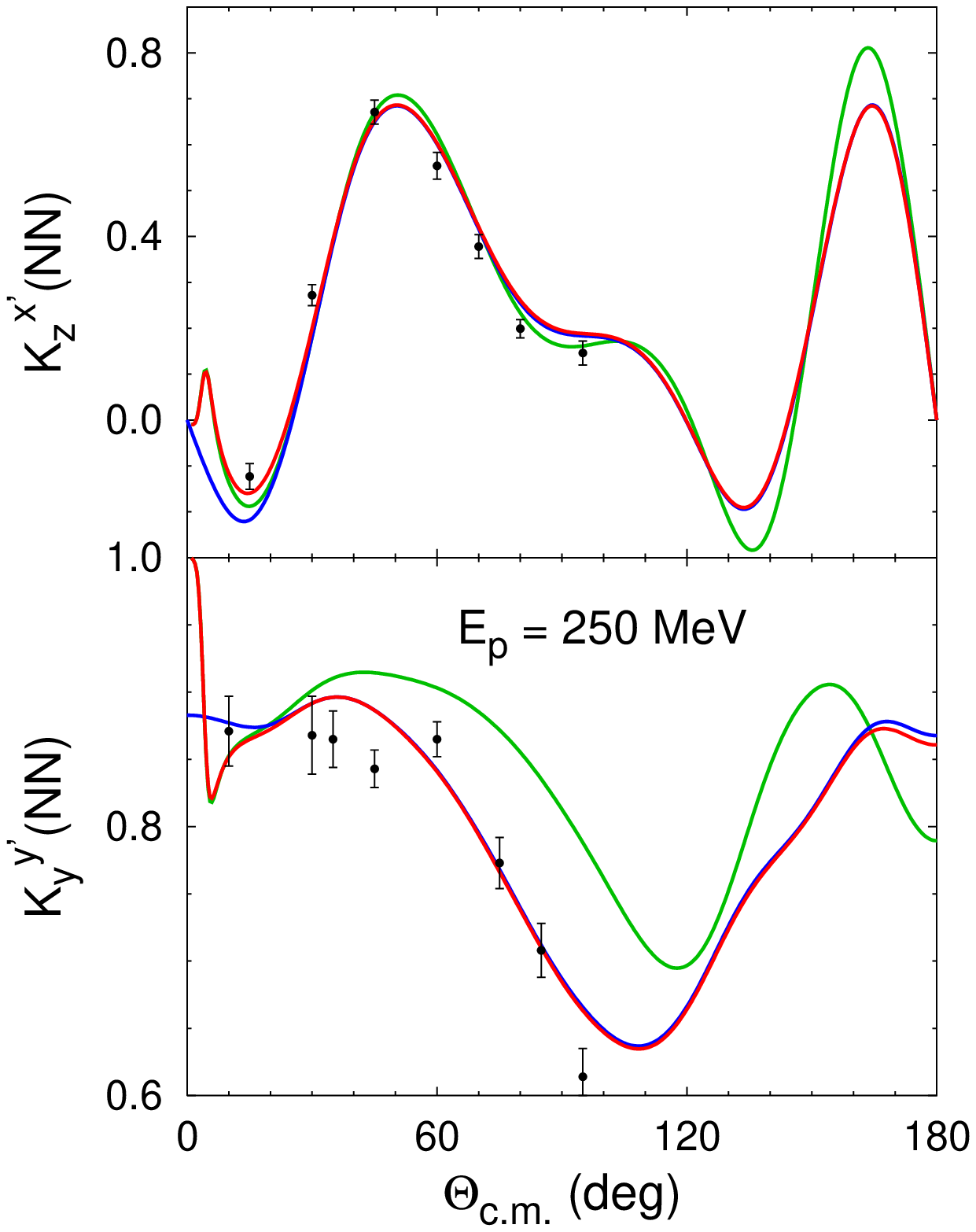}}
\caption{\label{fig:3NForce} Selected observables of pd elastic scattering at higher energies. The effect of the 3N force is quite pronounced. In contrast, an effect of the Coulomb repulsion between the protons is only seen in the extreme forward direction. At 135 MeV p energy, the experimental data are from Ref. \cite{ermisch:05}, the lower data in the differential cross section from Ref. \cite{sekiguchi:02}, an example for conflicting experimental data; the theoretical predictions are from Ref. \cite{deltuva:05a}. At 250 MeV p energy, the experimental data are from Ref. \cite{hatanaka:02}, the theoretical predictions from Ref. \cite{deltuva:11}.}
\end{figure}

The inclusion of the Coulomb repulsion between the two p's is necessary for the successful description of 3N and 4N elastic scattering at low energies. But Fig.~\ref{fig:Coulomb} shows that Coulomb can be quite important also at much higher beam energies, when,  in the breakup situation, the two outgoing p's are strongly correlated at rather low relative energies. Signals for the working of the 3N force in the considered dynamic model are shown in Fig.~\ref{fig:3NForce}. \\

\begin{figure}[!]
\centerline{\includegraphics[scale=0.45]{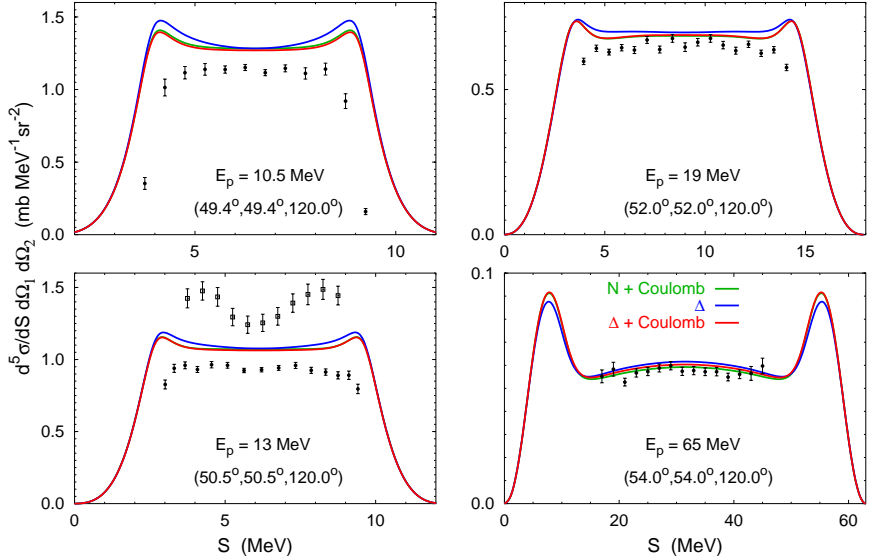}} 
\caption{\label{fig:BreakUp} Nd breakup in the space-star kinematics. Neither the Coulomb force between the two p's nor the 3N force show up in any significant way. The angles of two outgoing N's are fixed; their energies are constrained by the kinematical locus $\rm S$. The data refer to pd breakup with the exception of the upper data at 13 MeV N beam energy which are nd data. The theoretical predictions are from Ref. \cite{deltuva:05b} which also gives the references to the experimental data. }
\end{figure}

A very long-standing puzzle is the spin observable ${\rm A}_y$ in elastic pd, but also in elastic p${^3}\rm He$ scattering in a particular low-energy window. Another observable which is extremely hard to describe is the total elastic neutron-${{^3} \rm H}$ (${\rm n} {{^3} \rm H}$)  cross section. I like to discuss a further puzzle arising at low-energy pd breakup in the space-star kinematics. Data and the theoretical predictions are shown in Fig.~\ref{fig:BreakUp}. That space-star kinematics was believed by experimentalists to show the effect of the 3N force most strongly; in fact, that effect is not seen at all. At 13 MeV N lab energy there are pd and nd data; since the nd experiments are especially difficult, the data were twice remeasured, but appear now to be confirmed; the pd data were taken only once. There is a sizable difference between pd and nd data; theory is unable to account for that difference; the Coulomb effect is minor; if the data were true beyond any doubt, an extremely large nuclear charge-asymmetry effect shows up. Such an effect appears, however, conceptually rather unlikely. 

\subsection{Summary}
\begin{figure}[h]
\centerline{\includegraphics[width=0.85\textwidth]{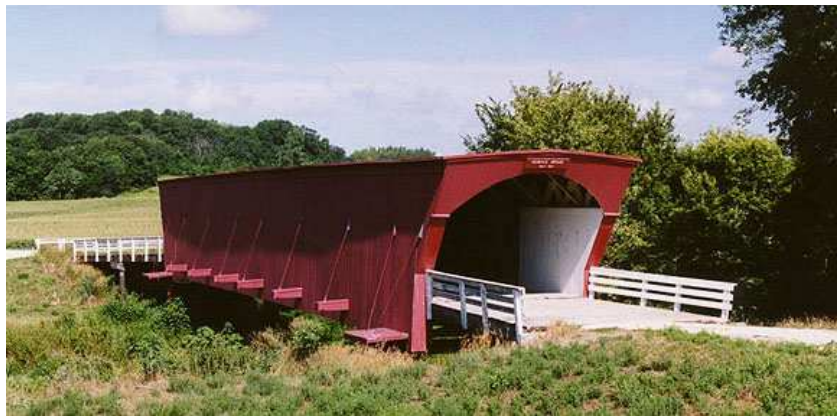}}
\caption{\label{fig:Bridge} One of the covered bridges of Iowa.}
\end{figure}

In the past, the theoretical fields of nuclear structure and few-nucleon systems were entirely disjoint with respect to research goals, to employed dynamics and to numerical techniques used for solving the nuclear many-body problems. Research settled on different banks of the river "nuclear theory". That situation passed; there are now interesting cross-overs between those fields as this conference in Iowa is witness for, and the beautiful bridges of Iowa as the one of Fig.~\ref{fig:Bridge} are pictures for those cross-overs. The talk discussed {\em genuine} and {\em effective} 2N and many-N forces, their appearance and their different roles in nuclear-structure and in few-nucleon calculations. The talk presented some examples for the achievements of few-nucleon theory, but also for outstanding puzzles in the description of data. \\

At the end, I wish the man of honor at this conference, James Vary, further success in his admirable engagement for the advancement of nuclear physics, which has been and will be stimulating to others. \\

The shown results for few-nucleon systems were obtained in a long successful collaboration with A. Deltuva and A.C. Fonseca, University of Lisbon, for which I am very grateful.

\end{document}